\pdfoutput=1 
\documentclass{JINST}

\usepackage{subfigure}
\usepackage{mathrsfs}
\usepackage{xspace}

\newcommand {\ie}{\mbox{i.e.}\xspace}     
\newcommand {\eg}{\mbox{e.g.}\xspace}     

\title{3D silicon pixel detectors for the ATLAS Forward Physics experiment}

\author{J. Lange$^a$\thanks{Corresponding author.}~,
E. Cavallaro$^a$,
S. Grinstein$^{a,b}$ 
and I. L\'{o}pez Paz$^a$\\
\llap{$^a$}Institut de F\'{i}sica d'Altes Energies (IFAE),\\
  08193 Bellaterra (Barcelona), Spain\\
\llap{$^b$}ICREA\\

E-mail: \email{jlange@ifae.es}}

\abstract{
The ATLAS Forward Physics (AFP) project plans to install 3D silicon pixel detectors about 210\,m away from the interaction point and very close to the beamline (2--3 mm). This implies the need of slim edges of about 100--200\,$\mu$m width for the sensor side facing the beam to minimise the dead area. Another challenge is an expected non-uniform irradiation of the pixel sensors. 

It is studied if these requirements can be met using slightly-modified FE-I4 3D pixel sensors from the ATLAS Insertable B-Layer production. AFP-compatible slim edges are obtained with a simple diamond-saw cut. Electrical characterisations and beam tests are carried out and no detrimental impact on the leakage current and hit efficiency is observed. For devices without a 3D guard ring a remaining insensitive edge of less than 15\,$\mu$m width is found. Moreover, 3D detectors are non-uniformly irradiated up to fluences of several 10$^{15}$~n$_{eq}$/cm$^2$ with either a focussed 23\,GeV proton beam or a 23\,MeV proton beam through holes in Al masks. The efficiency in the irradiated region is found to be similar to the one in the non-irradiated region and exceeds 97\% in case of favourable chip-parameter settings. Only in a narrow transition area at the edge of the hole in the Al mask, a significantly lower efficiency is seen. A follow-up study of this effect using arrays of small pad diodes for position-resolved dosimetry via the leakage current is carried out.}


\begin{document}

\section{Introduction}

ATLAS plans to install a new forward detector about 210\,m away from the interaction point (IP), the ATLAS Forward Physics detector (AFP)~\cite{bib:LetterOfIntentPhaseI}. The objective of this project is the detection of very forward protons (p) scattered under a small angle for the study of diffractive physics. The current AFP scenario foresees a low-luminosity operation during short dedicated LHC runs, whereas the system can be upgraded at a later stage to also take data at higher luminosities during large part of the regular LHC runs.

AFP is designed to include high-resolution tracking detectors in combination with fast timing detectors for event pile-up removal. For the tracking detectors, silicon pixel modules are foreseen based on the FE-I4 readout chip~\cite{bib:FEI4} as used in the recently installed ATLAS Insertable B-Layer (IBL)~\cite{bib:IBL3Dprod, bib:IBLprototypes}. However, in contrast to the IBL that surrounds the beam at the interaction point in a barrel configuration at a radius of 3.3\,cm, the AFP pixel modules will be placed almost perpendicular to the beam (under a small tilt of 15$^{\circ}$) with one side only 2--3\,mm away from it as it is vital for the physics program to measure very small scattering angles for a good acceptance~\cite{bib:LetterOfIntentPhaseI,bib:AFPTracker}. This leads to two additional requirements for the pixel detectors: 

\begin{enumerate}
\item The insensitive region of the detector side facing the beam has to be minimised to about 100--200\,$\mu$m. 
\item Due to the specific profile of scattered protons at the detector position, the devices have to withstand a highly non-uniform irradiation with a fluence difference of orders of magnitudes between different parts of the same sensor. The magnitude of the maximum fluence depends on the run scenario: about $5\times10^{12}$~p/cm$^2$ are expected for initial low-luminosity runs and about $5\times10^{15}$~p/cm$^2$ for a possible later high-luminosity scenario.
\end{enumerate}

Silicon pixel sensors based on the 3D technology are the baseline for AFP due to an excellent radiation hardness together with a low depletion voltage and their maturity proven by successful production runs for the IBL~\cite{bib:IBL3Dprod}. 
In the context of this study, it is investigated whether the IBL 3D sensors produced by FBK (Fondazione Bruno Kessler, Trento, Italy)~\cite{bib:FBKIBLProduction} and CNM (Centro Nacional de Microelectronica, Barcelona, Spain)~\cite{bib:CNMIBLProduction} with small modifications can also fulfil the additional AFP requirements. Different cutting techniques are studied to achieve AFP-compatible slim inactive edges. After first promising results~\cite{bib:AFP3D1} based on the advanced Scribe-Cleave-Passivate (SCP) technique, a simple diamond-saw cut is investigated here. Concerning radiation hardness, the IBL 3D sensors have proven to withstand a uniform fluence of 5$\times$10$^{15}$~n$_{eq}$/cm$^2$~\cite{bib:IBLprototypes}. However, in case of non-uniform irradiation like the one predicted for AFP, a scenario might occur in which the breakdown voltage of the non-irradiated region (which is usually lower than in irradiated silicon) is lower than the voltage needed to provide a sufficiently high electric field in the irradiated area for efficient charge collection. Thus, studies of non-uniform irradiations are performed. After first tests using a focussed 23~GeV p beam~\cite{bib:AFP3D1}, a more localised fluence deposition is achieved here using aluminium masks with holes in a 23~MeV p beam. The performance of the obtained slim-edge and non-uniformly irradiated devices is studied using electrical characterisations and test beams.

\section{Slim-edge AFP pixel modules \label{sec:AFPmodules}}

\begin{figure}[bt]
	\centering
	 \includegraphics[width=15cm]{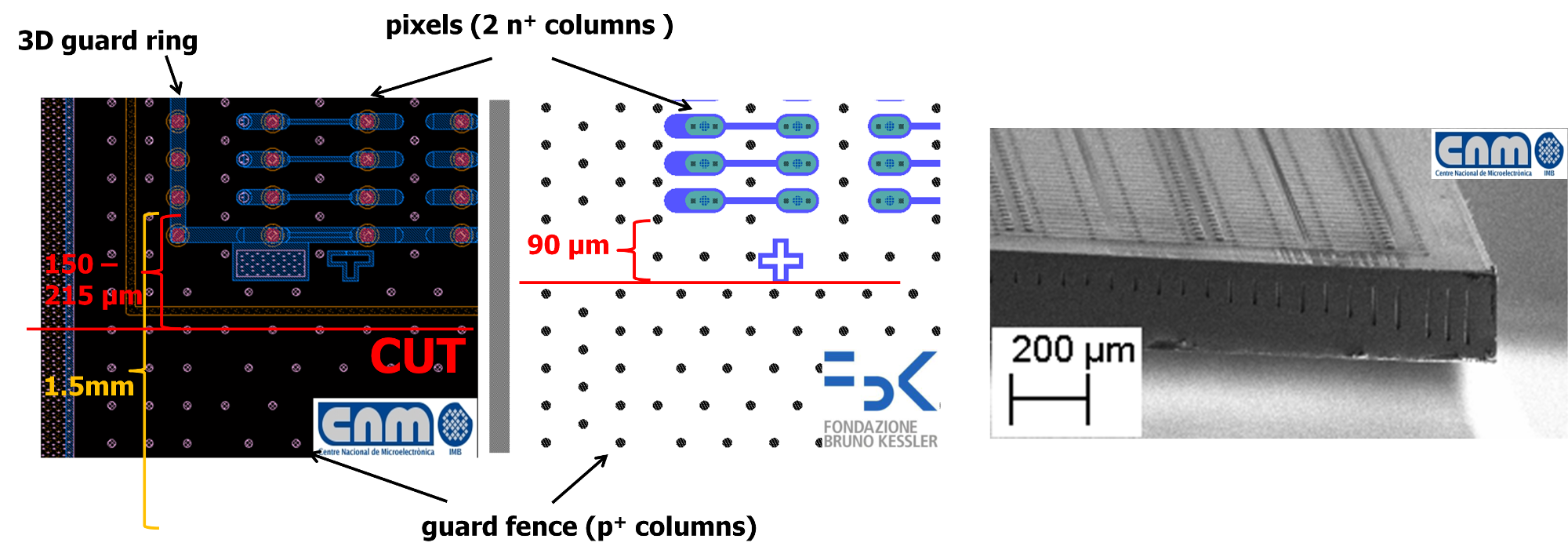}
	\caption{Sketch of a corner of a 3D CNM (left) and FBK (centre) pixel sensor. The cut lines for the AFP slim-edge prototypes are indicated. Picture of a CNM sensor after a diamond-saw cut (right).}
	\label{fig:slimEdgeSensor}
\end{figure}

The AFP prototypes based on IBL FE-I4 3D pixel detectors consist of an array of $80\times336$ pixels with $250\times50\,\mu$m$^2$ size each, giving an overall active area of $2.00 \times 1.68$\,cm$^2$. The 3D sensors fabricated by CNM and FBK in a double-sided process consist of a 230\,$\mu$m thick p-type sensitive material which is penetrated by the columnar 3D electrodes perpendicular to the surface~\cite{bib:IBL3Dprod,bib:FBKIBLProduction,bib:CNMIBLProduction}. Each pixel consists of 2 n$^+$-junction columns and 6 surrounding p$^+$-ohmic columns (see figure~\ref{fig:slimEdgeSensor}).

An important aspect with respect to the edge sensitivity is the edge termination of the pixelated area. CNM uses a 3D guard ring of n$^+$ columns that terminates the sensitive pixel area and drains leakage current from outside. It is surrounded by a p$^+$-column fence that terminates the depletion region growing from the guard ring and prevents it from reaching the side wall. In contrast, FBK only uses a p$^+$-column fence that terminates the depletion region growing from the last pixel. Nevertheless, that depletion region still can float into regions beyond the geometrical boundaries of the last pixels, thereby allowing an extension of the sensitive region (due to the absence of a guard ring). The designs of both CNM and FBK IBL 3D sensors already include a cut line only 200\,$\mu$m away from the two external pixel columns (left and right) in order to abut IBL sensors along that direction. The AFP-relevant bottom side opposite to the wirebonds, however, incorporates an insensitive 1.5\,mm wide region for a bias tab that is only needed for single-sided 3D processes, \ie not in the CNM and FBK cases. For the slim-edge AFP prototypes, large part of this is cut away with a diamond saw, leaving a remaining edge extension of about 90\,$\mu$m (FBK) and 150 to 215\,$\mu$m (CNM) as shown in figure~\ref{fig:slimEdgeSensor} and table~\ref{tab:slimEdges}. Such a standard diamond-saw cut has the advantage compared to more advanced methods like Scribe-Cleave-Passivate, which was studied before~\cite{bib:AFP3D1}, that it can be performed directly and fast at CNM and FBK. 

It should be noted that these devices were built from spares of the IBL production, thereby not always fulfilling the IBL quality criteria regarding \eg leakage current and breakdown voltage.

\begin{table}[tbh]
\caption{Overview on produced slim-edge AFP 3D pixel detectors and measured properties of the slimmed bottom edge at 20\,V (FBK) and 30\,V (CNM).}
	\centering
	\small
		\begin{tabular}{|l|r|r|r|r|}
		\hline
		Sample 	                                & FBK-S1-R9 & FBK-S2-R10 & CNM-S3-R5   & CNM-S5-R7	\\
		Edge extension after cut                & 91\,$\mu$m& 87\,$\mu$m & 215\,$\mu$m & 150\,$\mu$m\\
		Sensitivity extension beyond last pixel & 77\,$\mu$m& 75\,$\mu$m &   1\,$\mu$m & 7\,$\mu$m \\
		Remaining insensitive edge              & 14\,$\mu$m& 12\,$\mu$m & 214\,$\mu$m & 143\,$\mu$m \\
		\hline
		\end{tabular}
	
	\label{tab:slimEdges}
\end{table}

\section{Performance of slim-edge AFP pixel modules  \label{sec:slimEdgePerformance}}

\begin{figure}[b]
	\centering
	\includegraphics[width=10cm]{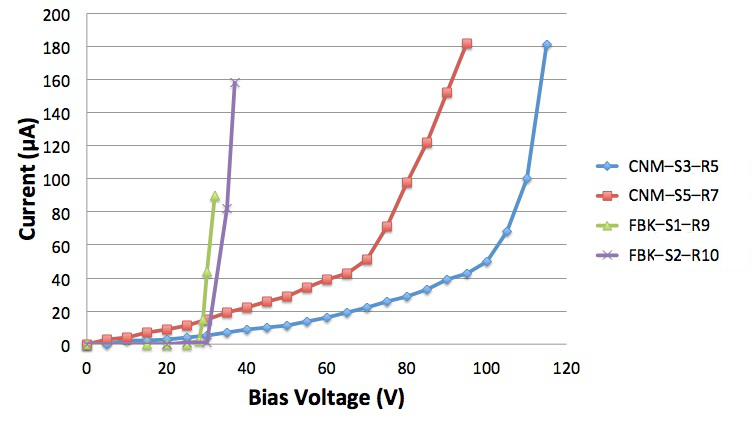}
	\caption{Current-voltage characteristics for the slim-edge AFP protoypes.}
	\label{fig:AFP_FEI4_IV}
\end{figure}

It needs to be tested whether a simple diamond-saw cut without side-wall passivation does not introduce too much leakage current and that the hit efficiency remains high, especially in the edge region. 
Figure~\ref{fig:AFP_FEI4_IV} shows the current-voltage (IV) characteristics of the studied slim-edge AFP sensors after bump-bonding and assembly to a flexible circuit board. The FBK sensors exhibit a low current below 1\,$\mu$A up to a sharp breakdown voltage of about 30\,V. On the contrary, the CNM sensors show a steady current increase up to about 50\,$\mu$A, still below the breakdown voltages of 70 to 100\,V. Such curves are typical of the IBL-spare quality class and do not show any abnormal behaviour that could be attributed to the edge slimming. This is consistent with a previous study on FBK sensors which shows that no strong current increase occurs up to a diamond-saw cut at about 75\,$\mu$m distance to the edge pixels~\cite{FBKcutStudy} (unfortunately, for CNM sensors it is not possible to obtain IV curves for the whole sensor before bump-bonding).

The measured mean noise of about 150--160\,e$^-$ for the CNM devices and 200\,e$^-$ for the FBK ones is a bit higher than the IBL average of 130--140\,e$^-$~\cite{bib:IBLprototypes}, but typical of the IBL-spare quality class available here. 
Most of the devices show no excess noise at the edges. Only CNM-S5-R7 exhibits up to 40\% higher noise in the pixel rows close to the slimmed edge, which however has no impact on the operation with a standard threshold of 2\,ke$^-$.

\begin{figure}[bt]
	\centering
	\includegraphics[width=15cm]{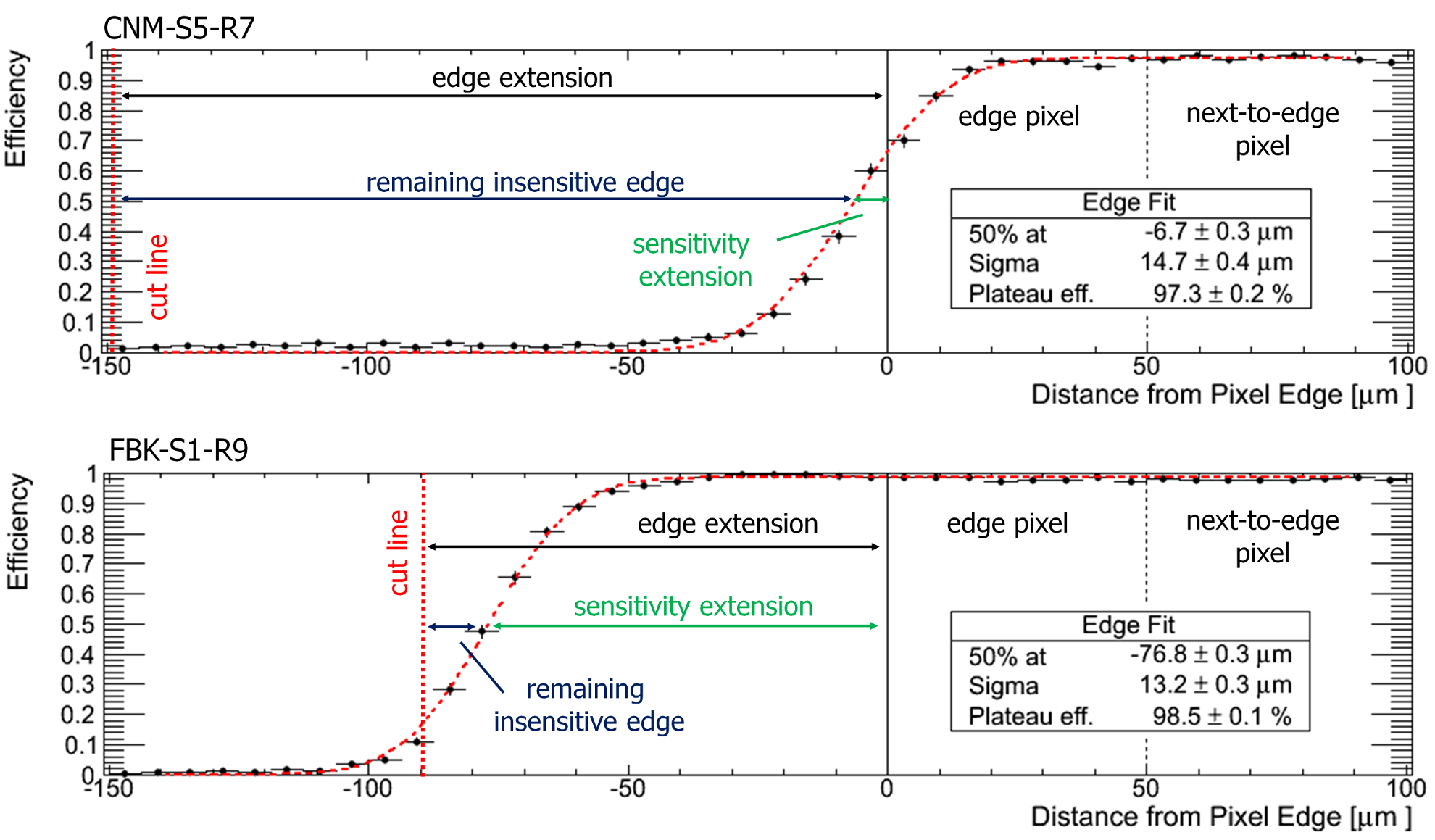}
	\caption{Efficiency in the region of the AFP-relevant bottom edge as a function of distance from the edge of the last pixel row for CNM-S5-R7 (top) and FBK-S1-R9 (bottom).}
	\label{fig:edgeEff}
\end{figure}

The hit efficiency is determined in test beams at DESY using 4 or 5\,GeV electrons. The ACONITE telescope based on EUDET~\cite{bib:EUDET}, is used to extrapolate beam tracks in the devices under test with a precision of about 15\,$\mu$m at the electron energies used. Two devices under test (each time one of the four AFP prototypes and a regular-edge IBL CNM reference sensor) are placed simultaneously under a perpendicular beam incidence angle between the telescope planes. Bias voltages of 20\,V (FBK) or 30\,V (CNM) are applied and the threshold of the FE-I4 readout chip is tuned to 2--3\,ke$^-$. The overall efficiencies of the AFP prototypes are measured to be 97--99\%, similar to the one of the reference sensor (98--99\%). The efficiency around the edge of the last pixel row at the slimmed bottom side of the sensor is shown in figure~\ref{fig:edgeEff} for one CNM and one FBK AFP prototype. Similar results are obtained for the second CNM and FBK devices as well as for the edge efficiency around the non-slimmed top side of the sensors. The efficiency is found to be stable up to the last pixel row for CNM devices (the smearing of the step is due to the telescope pointing resolution), and even beyond in case of the FBK sensors due to the absence of a 3D guard ring as discussed in section~\ref{sec:AFPmodules}. Table~\ref{tab:slimEdges} includes the width of the sensitive region beyond the edge of the last pixel row as obtained from the 50\% efficiency point. The sensitivity extension of about 75\,$\mu$m for the FBK sensors implies a remaining insensitive edge of less than 15\,$\mu$m, which is to the knowledge of the authors the slimmest edge measured in pixel detectors apart from sensors with fully active edges~\cite{bib:activeEdge1, bib:activeEdge}. However, it has to be considered that such a behaviour, which is beneficial from the efficiency point-of-view, implies a degradation of the position resolution of the last row.  
Also the CNM sensors with insensitive edges of 143--214\,$\mu$m fulfil the AFP requirements.

\section{Non-uniform irradiation of AFP pixel modules \label{sec:nonUniformIrrad}}

The studies of non-uniform irradiations performed here are targeted at the harsh requirements of the high-luminosity AFP run scenario with an expected total integrated luminosity of about 100\,fb$^{-1}$. At the position of the AFP pixel modules, it is expected that the total accumulated fluence of scattered beam protons with energies up to slightly below 7\,TeV reaches a maximum of about 5$\times$10$^{15}$~p/cm$^2$ along a few mm wide and about 1\,cm long line that starts from the beam (this shape is due to the LHC magnet configuration between the IP and AFP)~\cite{bib:AFPTracker}. Already very close to that line, the fluence is expected to be orders of magnitude lower within the same pixel detector. The contributions from beam backgrounds have not yet been simulated, but are expected to add a more uniform fluence distribution from lower-energy particles. For the planned initial low-luminosity run scenario, a 3 orders of magnitude lower integrated luminosity and hence also fluence is expected.

\begin{figure}[htbp]
	\centering
	 \includegraphics[width=15cm]{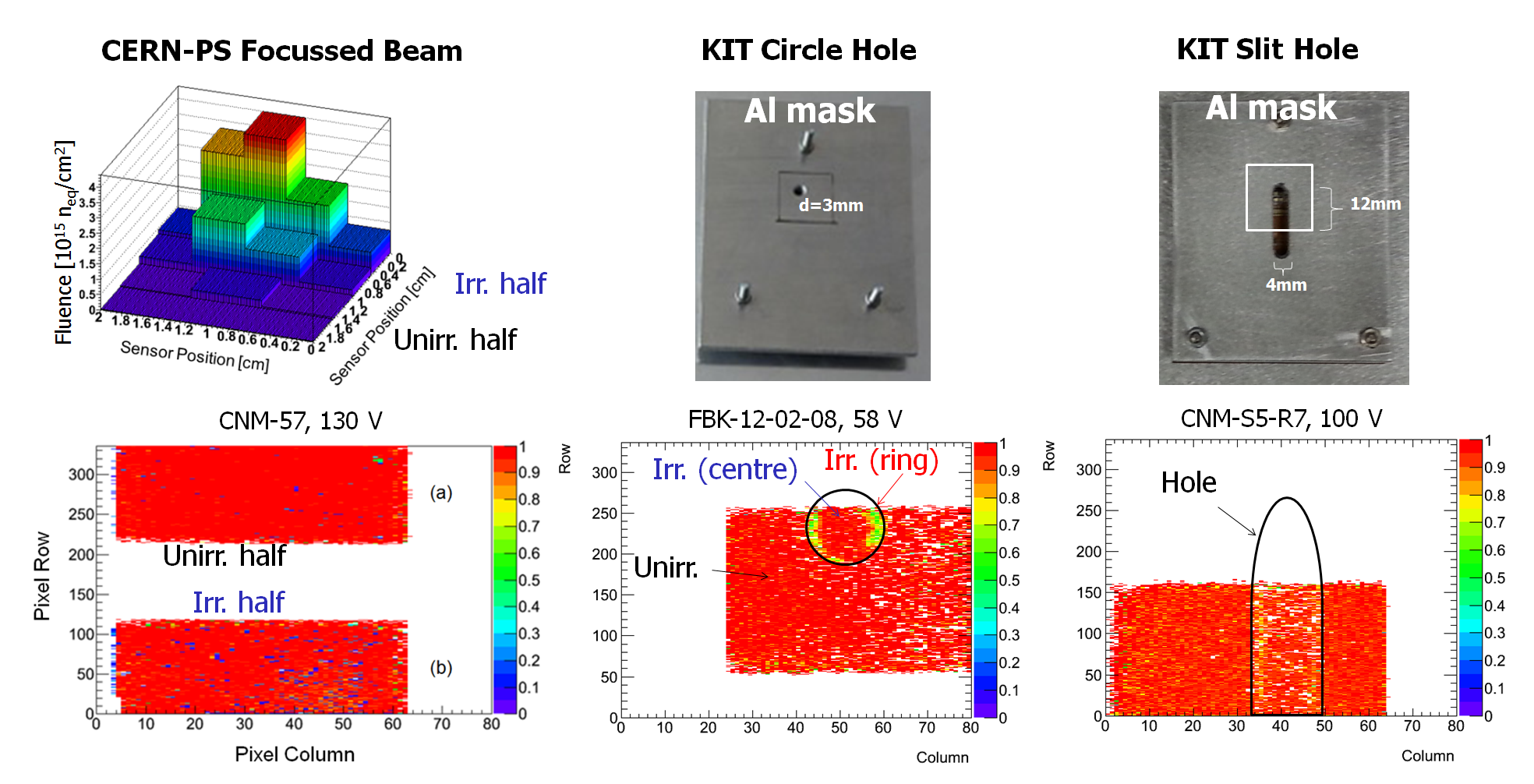}
	\caption{Irradiation mode (top) and measured efficiency maps (bottom). The efficiency is only measured for a part of the sensor since the smaller telescope planes give reference tracks only for that area (the map for CNM-57 consists of two separate measurements). The CERN-PS plots are reproduced from~\cite{bib:AFP3D1}.}
	\label{fig:nonUniformEff}
\end{figure}

In absence of a multi-TeV p irradiation facility, proof-of-principle non-uniform-irradiation experiments at existing facilities of lower p energies have been performed. It should be noted that even theoretically the p damage in silicon is only well studied up to 23\,GeV~\cite{bib:Huhtinen}. Assuming a similar hardness factor of $\kappa\approx0.6$ at TeV as at GeV p energies, the expected maximum proton fluence for the high-luminosity running scenario corresponds to a 1~MeV n equivalent fluence of 3$\times$10$^{15}$\,n$_{eq}$/cm$^2$. Irradiation campaigns have been performed at two different facilities with different degrees of non-uniformities:

\begin{enumerate}

\item A focussed 23\,GeV p beam at CERN-PS is used to give a maximum equivalent fluence of 4$\times$10$^{15}$\,n$_{eq}$/cm$^2$ (see \cite{bib:AFP3D1}). The fluence spread is relatively large (a region of more than 1\,cm diameter received a fluence of more than 10$^{15}$\,n$_{eq}$/cm$^2$ and even peripheral pixels acquired a fluence of 10$^{13}$--10$^{14}$\,n$_{eq}$/cm$^2$, see figure~\ref{fig:nonUniformEff}, top left). 

\item A more localised irradiation with an abrupt transition between irradiated and unirradiated area is achieved by using 5\,mm thick Al masks which can shield 23\,MeV p at KIT (Proton irradiation facility at the Karlsruhe Institute of Technology, Karlsruhe, Germany) and let them pass only through a hole. Either a circular hole of 3\,mm diameter is used or a slit-like hole of 4\,mm width and 12\,mm effective length over the sensor with fluences of 2--3.6$\times$10$^{15}$\,n$_{eq}$/cm$^2$ (see figure~\ref{fig:nonUniformEff}, top centre and top right).

\end{enumerate}

\begin{table}[htbp]
\caption{Samples and irradiation details of the non-uniformly irradiated devices (top), parameters during the test beam measurements (centre) and measured hit efficiency for different sensor regions (bottom).}
	\centering
	\small
		\begin{tabular}{|l|c|c|c|c|c|}
		\hline
		\multicolumn{6}{|c|}{Device and Irradiation Details}	\\
		\hline
		Irradiation type 	& Unirr. ref. 	 &     PS focussed 	& KIT	circ. hole         & \multicolumn{2}{|c|}{KIT slit hole}	\\
		\hline
		Fluence [10$^{15}$\,n$_{eq}$/cm$^2$]& 	Unirr. &	4.0 (max) &	1.8	&3.3	&3.6 \\
		\hline
		Sample &	CNM-55	& CNM-57 &	FBK-12-02-08 & CNM-S5-R7 &	CNM-S3-R5 \\
		\hline
		\hline
		\multicolumn{6}{|c|}{Measurement and chip parameters}	\\
		\hline
		Threshold [ke$^-$] &	3 &	1.7 &	2 &	2 &	3 \\
		\hline
		ToT at 20\,ke$^-$ &	10	& 10	& $\approx$11&	$\approx$5 &	$\approx$8 \\
		\hline
		Single small hits rejected &	No &	No &	No &	Yes &	Yes \\
		\hline
		V$_{bias}$ [$V$]        & 30& 130&  58 & 100 & 130 \\ 
		\hline
		\hline
		\multicolumn{6}{|c|}{Measured hit efficiency [\%]}	\\
		\hline
		Unirr. region	    &99 &	99	& 98 &	95 &	94 \\
		\hline
		Irr. region (centre) &	- &	98	& 97 &	94 &	93 \\
		\hline
		Irr. region (ring) 	& -	& -	  & 70 &	90 &	58 \\
		\hline

		\end{tabular}
	
	\label{tab:TestBeamResults}
\end{table}

In the second scenario the non-uniformity is apparently more extreme and resembles more the expectation when considering only the fluence from the scattered beam protons. However, that line of scattered p might be smeared a bit during operation (\eg due to movements of the detectors or changes in the LHC magnet parameters) and beam background is not yet taken into account in the simulation. Hence in the end, a non-uniformity between the first and second scenario might be realistic. The samples used, together with the level of irradiation, are listed in table~\ref{tab:TestBeamResults}. CNM-S5-R7 and CNM-S3-R5 already include the slim edge as discussed in the previous section.


The performance of these samples is again measured in test beams at DESY with 4 or 5\,GeV electrons or at CERN-SPS with 120\,GeV pions at temperatures between -15 and -50$^\circ$C to reduce the leakage current (dry-ice cooling). Table~\ref{tab:TestBeamResults} shows an overview on the tuning and parameters of the FE-I4 chip and the resulting efficiencies for different regions of the sensor at the maximum applied voltage (limited by leakage current and/or a noise increase). The different regions are also well visible in the measured efficiency sensor maps as shown in figure~\ref{fig:nonUniformEff}. The sensor irradiated with the focussed PS beam is only subdivided into the irradiated side (with the maximum fluence) and the unirradiated side (to be precise: the side that received a low fluence). In contrast, for the samples irradiated through the Al mask at KIT, it can be distinguished between the unirradiated region, the centre of the irradiated region and a ring around the edge of the hole. Almost full efficiency of 98--99\% is obtained in the unirradiated region of CNM-57 and FBK-12-02-08, similar to the unirradiated reference sensor. The overall lower efficiency for CNM-S5-R7 and CNM-S3-R5 is explained by an unfavourable setting in the FE-I4 chip (HitDiscConfig=2) that leads to the rejection of single small hits (\ie with time-over-threshold $ToT<3$). In a recent test beam CNM-S5-R7 has been remeasured with more favourable settings, and in a preliminary analysis a higher efficiency similar to the other devices is confirmed. The efficiency in the irradiated region (only the centre in case of the masks) approaches within 1\% the one in the unirradiated area in all cases. However, there is a ring of lower efficiencies around the hole at KIT irradiations. The values there vary substantially between 58--90\%.


The reason for the low-efficiency ring is still under investigation. It is conceivable that the edge region might have obtained a higher equivalent fluence due to scattering of p at the edge of the Al hole, thereby loosing energy, which leads to an increase in their displacement damage cross section in Si~\cite{bib:Huhtinen}. Alternatively, it might be a real sensor effect in situations with such an abrupt transition between irradiated and unirradiated regions within one single silicon sensor, leading to a large gradient in defect density and leakage current. Or it might be an effect of the readout chip.

To check the first hypothesis, another irradiation is performed at KIT with the same Al masks with the slit-like hole to 3.5$\times$10$^{15}$\,n$_{eq}$/cm$^2$, this time with a $4\times4$ array of small circular p-on-n silicon pad diodes (0.5\,mm diameter, 1.5\,mm pitch, 300\,$\mu$m thickness). A position-resolved dosimetry is obtained from measuring the leakage current (at 20$^\circ$C and after annealing for 80\,min at 60$^\circ$C) and using the linear relation between the current and the fluence with the damage parameter $\alpha=4\times$10$^{-17}$\,A/cm~\cite{bib:Moll}. At this high fluence, no real plateau is observed in the IV curves, but at about 400\,V the slope is decreasing and the fluence values evaluated at 400\,V are consistent with the nominal fluence reported by KIT. No difference is found between the centre of the hole and the edge, thereby excluding a higher equivalent fluence (at least as obtained from the leakage current). Charge measurements with $^{90}$Sr of the different diodes are ongoing to check whether the edge of the hole has an influence on the charge collection. Further irradiation tests and simulations are envisaged to study whether the low-efficiency ring is a sensor or a chip effect.

\section{Summary and conclusions \label{sec:conclusions}}
Slim-edge and non-uniformly irradiated 3D pixel detectors from FBK and CNM are studied with electrical characterisations and beam tests in view of applications at forward detectors like AFP.

It is shown that it is possible to reduce the unpixelated edge extension opposite the wirebonds with a simple diamond-saw cut down to 87--215\,$\mu$m width without compromising the leakage current, noise and hit efficiency. The FBK design without a 3D guard ring even leads to efficient regions up to 77\,$\mu$m beyond the edge of the last pixel row, thereby further reducing the insensitive area to less than 15\,$\mu$m, which is only surpassed by sensors with fully active edges to date.

3D detectors are non-uniformly irradiated with either a focussed beam at CERN-PS or holes in Al masks at KIT to fluences expected after the high-luminosity run scenario at AFP (several 10$^{15}$\,n$_{eq}$/cm$^2$). High efficiencies of at least 97\% both in the unirradiated and irradiated parts are obtained (for favourable chip-parameter settings). Only in a narrow ring around the edge of the hole for irradiations with the Al mask a lower efficiency is observed. Leakage-current measurements in non-uniformly irradiated arrays of small diodes do not show any hints of a higher fluence in the edge region. Further investigations are needed to clarify its origin. However, the first approved low-luminosity AFP program requires only a 3 orders of magnitude lower maximum fluence and, furthermore, due to beam backgrounds and relative movements of the particle profile, such an abrupt fluence transition is expected to be avoided.

To conclude, slim-edge 3D pixel detectors have demonstrated their ability to fulfil the requirements of forward detectors very close to the beam like AFP.

\acknowledgments

We thank our colleagues from CNM (G.\,Pellegrini, M.\,Baselga, V.\,Greco) and FBK (G.-F.\,Dalla Betta) for providing samples and facilities and fruitful discussions. We are grateful to the AFP and RD50 collaborations for many useful hints and discussions. We thank M.\,Glaser (CERN-PS) and F.\,B\"{o}gelspacher (KIT) for help with the irradiations and I.\,Rubinsky, A.\,Micelli, D.\,Pohl, O.\,Korchak and Sh.\,Hsu for help during the testbeams. This work was partially funded by the MINECO, Spanish Government, under grants FPA2010-22060-C02, FPA2013-48308-C2-1-P and SEV-2012-0234 (Severo Ochoa excellence program) and the European Commission under the FP7 Research Infrastructures project AIDA, grant agreement no. 262025.

\end{document}